\documentclass[structabstract]{aa}  
\usepackage{graphicx}
\usepackage{natbib}
\usepackage{color}
\usepackage{url}
\usepackage{txfonts}

\newcommand{\RSUN}{R$_\odot$ }
\newcommand{\jasr}{    {\it Adv. Space Res.}}
\begin{document}
   \title{The relativistic solar particle event of 2005 January 20: \\ 
    origin of delayed particle acceleration}

   \titlerunning{The GLE of 2005 January 20: origin of delayed particle acceleration}


   \author{K.-L. Klein
          \inst{1}
          \and
          S. Masson \inst{1,} \inst{2,} \inst{3}
          \and
          C. Bouratzis\inst{4}
           \and
           V. Grechnev\inst{5}
           \and
          A. Hillaris\inst{4}
          \and
          P. Preka-Papadema\inst{4}
         }

   \institute{
   	LESIA-UMR 8109, Observatoire de Paris, CNRS, Univ. Paris 6 \& 7,
              Observatoire de Meudon, F-92195 Meudon, France; 
              \email{ludwig.klein@obspm.fr}
             \and
 	      Space Weather Laboratory, 
					NASA-Goddard Space Flight Center, 
					8800 Greenbelt Road, Greenbelt, MD 20771, USA;
             \email{sophie.masson@nasa.gov}
             \and
             Catholic University of America, 620 Michigan Ave NE, Washington DC, 20064, USA, 
              \and
              Section of Astrophysics, Astronomy and Mechanics, Department of Physics,
		University of Athens, Zografos (Athens), GR-15783, Greece; 
              \email{kbouratz@phys.uoa.gr}
             \and
             Institute of Solar-Terrestrial Physics SB RAS, Lermontov St. 126A, Irkutsk 664033, Russia; 
             \email{grechnev@iszf.irk.ru}
             }

   \date{Received 10/03/2014; accepted 14/08/2014}
 
  \abstract
   {  
   The highest energies of solar energetic nucleons detected in space or through gamma-ray emission in the solar atmosphere are in the GeV range. Where and how the particles are accelerated is still controversial.
   }
   {
   We search for observational information on the location and nature of the acceleration region(s) by comparing the timing of relativistic protons detected on Earth and radiative signatures in the solar atmosphere during the particularly well-observed 2005 Jan 20 event. 
    }
   {
   This investigation focusses on the post-impulsive flare phase, where a second peak was observed in the relativistic proton time profile by neutron monitors. This time profile is compared in detail with UV imaging and radio spectrography over a broad frequency band from the low corona to interplanetary space.
   } 
   {It is shown that the late relativistic proton release to interplanetary space was accompanied by a distinct new episode of energy release and electron acceleration in the corona traced by the radio emission and by brightenings of UV kernels. These signatures are interpreted in terms of magnetic restructuring in the corona after the CME passage.  
 }
   {We attribute the delayed relativistic proton acceleration to magnetic reconnection and possibly to turbulence in large-scale coronal loops. While Type II radio emission was observed in the high corona, no evidence of a temporal relationship with the relativistic proton acceleration was found.}

   \maketitle
%

\section{Introduction}
\label{Sec_Intro}
On certain occasions transient energetic particle fluxes from the Sun, called solar energetic particle (SEP) events, may comprise relativistic nucleons at energies up to several GeV or even tens of GeV. Upon impinging on the Earth's atmosphere, these particles trigger nuclear cascades that produce secondaries detectable by ground-based neutron monitors and muon telescopes. These particular SEP events are therefore called ground level enhancements (or ground level events; GLEs). The rarity of GLEs - only 72 were detected since 1942, including a small event on 2014 Jan 06 - clearly shows that GeV energies are extreme in solar events. Understanding their origin is therefore one of the more challenging tasks in the research on solar eruptive activity. A comprehensive summary of GLE observations was given by \cite{Lop-06}, and the review by \cite{Car-62} is still very informative.

GLEs are produced in conjunction with intense flares and extremely fast coronal mass ejections \citep[CMEs;][]{Blv:al-10,Gop:al-12}. The acceleration mechanisms are thought to be related to the flare, which usually means magnetic reconnection, or to the shock wave generated by the CME. Which of the two possibilities, the flare or the shock wave, is actually the accelerator is hard to say on observational grounds. Physical relationships between the particles detected on Earth and in dynamical processes in the solar corona can in principle be inferred from comparing the arrival times of the SEPs and radiative signatures of flares and CMEs. This is especially possible when the SEP time profile has an intrinsic structure.

Evidence of successive distinguishable SEP releases within a given event has been reported from particle observations in the MeV to tens-of-MeV range \citep[e.g.,][]{Koc:al-07} and more often in GLEs, because the scattering mean free path in the interplanetary medium is larger at relativistic energies \citep{Dro-00}. It is indeed well established that GLEs often have a double-peaked structure, with an initial fast rise and an anisotropic particle population - called the `prompt component' - followed by a more gradual and less anisotropic `delayed component' \citep[see review in ][chap. 7.3]{Mir-01}. \cite{McC:al-12} conclude that the sequence of an anisotropic impulsive peak, and a less anisotropic gradual peak occurring 7-15~min later is a common occurrence when the parent active region is magnetically connected to the Earth, while the absence of the impulsive peak is typical of poorly connected activity near to or east of the central solar meridian or well beyond the western limb.

A prominent case illustrating this double-peaked structure is the GLE of 2005 Jan 20. It displayed a well-defined, rapidly rising time profile at the beginning and a distinct second peak a few minutes later. Evidence that the first release was related to particle acceleration in the flaring active region in the low corona was brought by different publications \citep[e.g.,][]{Sim-06,Sim-07, Kuz:al-08,Grc:al-08,McC:al-08,Msn:al-09}. In the present paper we complete the investigation of \cite{Msn:al-09}\footnote{\cite{Msn:al-09} will be cited as Paper~1 in the following.} through a detailed comparison between the second peak of the relativistic proton time profile derived from neutron monitor measurements with high-quality radio and UV observations. 

This article is structured as follows. Section~\ref{Sec_Obs} introduces the observations (\ref{Sec_Obs_inst}) and describes the time profile of relativistic protons detected at Earth (\ref{Sec_Obs_p}). We conclude that two successive proton releases near the Sun can be identified in the GLE observation. The evolution of radio, hard X-ray, and gamma-ray emissions is tentatively separated into different acceleration episodes in Sect.~\ref{Sec_Obs_eps}, and the relation with relativistic protons before and during the first release is briefly discussed in Sect.~\ref{Sec_Obs_pr1} with reference to Paper~1 and to the evolution of the dm-m-wave dynamic spectrum studied by \cite{Brt:al-10}. A more detailed analysis of the radio and UV emission accompanying the second relativistic proton release (Sect.~\ref{Sec_Obs_pr2}) leads us to suggest that the particles are accelerated during dynamical processes in the magnetically stressed corona after the CME passage. The findings are summarised in Sect.~\ref{Sec_Obs_sum} and discussed in Sect.~\ref{Sec_Disc} with respect to previous work on the origin of relativistic solar nucleons. 

   \begin{figure}
   \centering
   \includegraphics[width=9cm]{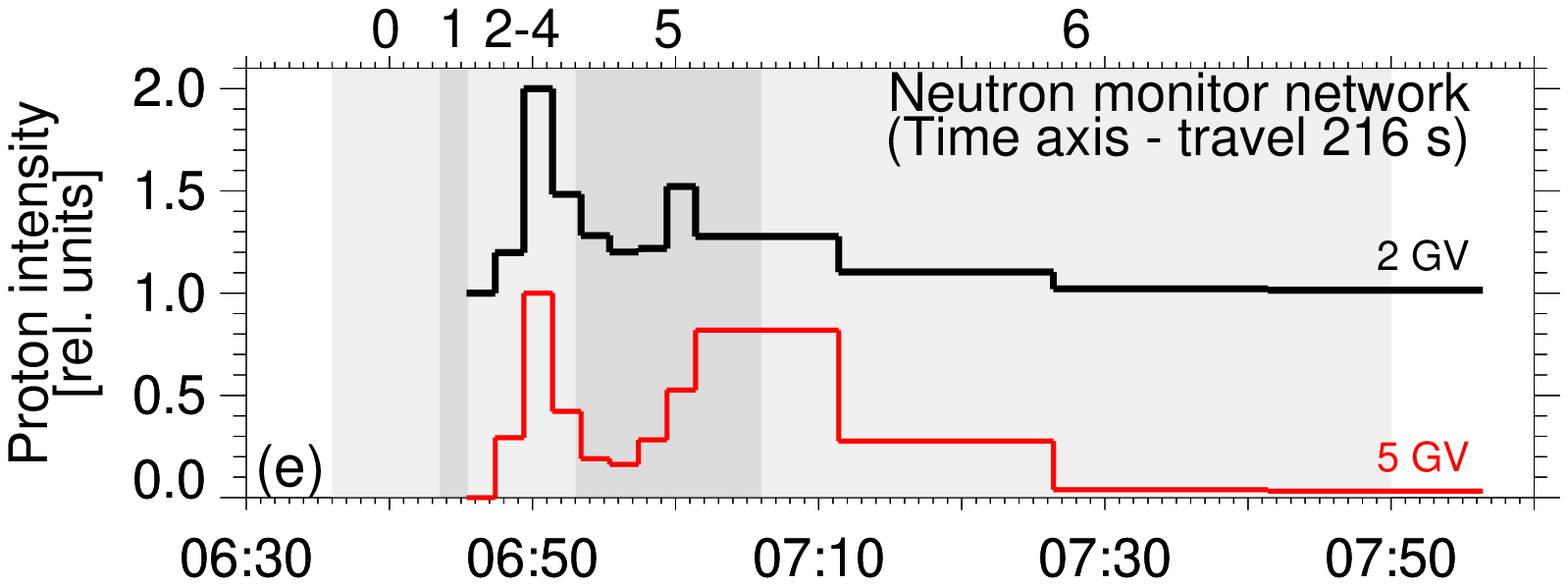}
   \includegraphics[width=9cm]{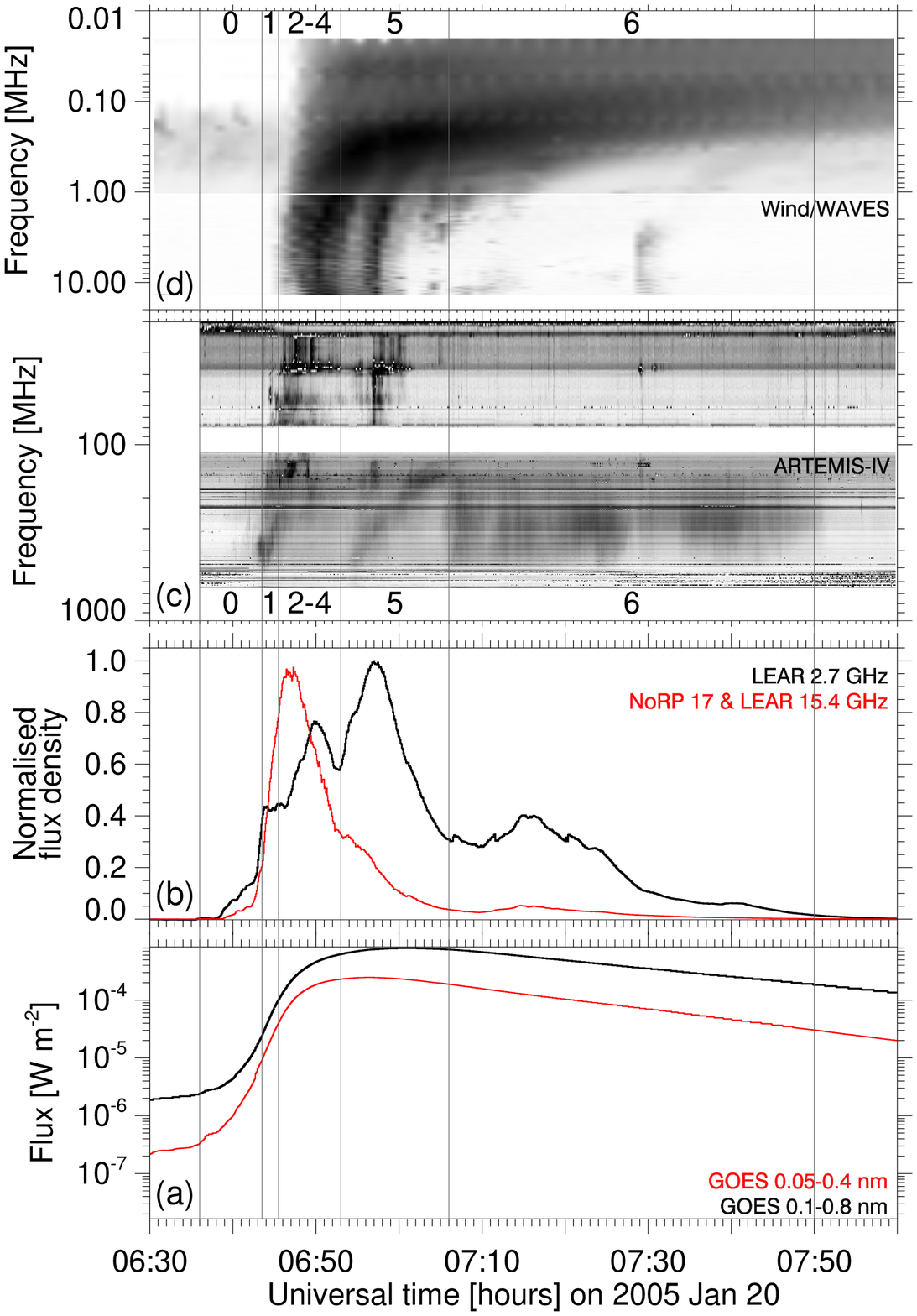}
     \caption[]{X-ray and radio emission and the relativistic proton profile of the 2005 Jan 20 event. From bottom to top: 
     (a) soft X-rays $\lambda=0.1-0.8$~nm (dark line) and 0.05-0.4~nm (light; red in the colour plot of the online version); 
     (b) microwaves (dark line 2.7 GHz, light  - red in the colour display -  a combination of 17 GHz (NoRP) before and 15.4~GHz (LEAR) after 06:55 UT); 
     (c) dynamic radio spectrum at dm-m waves (ARTEMIS-IV; inverse colour scale; 1~s integration time);
     (d) decametre-kilometre wave radio emission (Wind/WAVES; inverse colour scale; 1~min integration);
     (e) proton flux time history at 2~GV (dark curve) and 5~GV (light curve; red in the online version) rigidity (kinetic energy 1.27 and 4.15~GeV, respectivelyy), time axis shifted back by 216~s. 
     The intervals delimited by vertical lines and numbered 0 to 6 are different episodes of particle acceleration, as discussed in the text.
     }
  \label{Fig_ovw}
   \end{figure}


\section{Relativistic proton releases and electromagnetic emission during the 2005 Jan 20 flare and CME}
\label{Sec_Obs}

Relativistic protons penetrating the Earth's atmosphere are observed by the worldwide network of neutron monitors. Since individual neutron monitors detect signals from protons with different arrival directions and different low-energy limits, depending on the instrument's location within the Earth's magnetic field, the observations of the network can be inverted using a fitting procedure, and parameters of the energy spectrum (actually the rigidity spectrum) and the angular distribution of the arriving protons can be derived. This was done by  \citet{Bue:al-06}, using a power-law spectrum in magnetic rigidity. The results are also described in  Paper~1. Time histories of the proton intensities at rigidities of 2 and 5 ~GV, corresponding respectively to kinetic energies of 1.27 and 4.15~GeV, are displayed in the top panel of Fig.~\ref{Fig_ovw} (Fig.~\ref{Fig_ovw}.e). 

\subsection{ X-ray and radio emission}
\label{Sec_Obs_inst}

To elucidate the acceleration processes, we compared the relativistic proton time profiles with electromagnetic tracers of energy release, electron acceleration, and electron propagation from the corona to 1~AU. Whole Sun records in the X-ray and radio band are displayed in the lower panels of Fig.~\ref{Fig_ovw}  (from bottom to top):
\begin{itemize}
\item[(a)] soft X-ray (SXR) fluxes tracing the evolution of the plasma heated during the flare;
 \item[(b)] flux density time profiles at selected  microwave radio frequencies ($\nu \geq 1$~GHz), which is gyro-synchrotron radiation of near-relativistic electrons (energies between about 100~keV and a few MeV);
 \item[(c-d)] decimetre-to-kilometre wave dynamic spectrograms. While gyro-synchrotron emission may contribute at dm-m wavelengths, showing up as a more or less uniform background of moderate flux density in the dynamic spectrogram, the emission processes are mostly collective, leading to radio waves near the electron plasma frequency and its harmonic. The structuring in the frequency-time plane reveals non-thermal electrons with an ordered motion through the corona, such as electron beams (Type~III bursts) and shock waves (Type~II bursts), or which are confined in quasi-static or expanding magnetic structures (Type~IV bursts).
 \end{itemize}
 The ordering of the panels in Fig.~\ref{Fig_ovw} illustrates energy release and electron acceleration from the low corona (SXR and microwaves from loop structures extending to a few $10^4$~kms above the photosphere) to 1~AU (radio waves at a few tens of kHz). Roughly speaking, the radio emitting frequency decreases with increasing height above the photosphere. The reader is referred to the reviews of \cite{Bas:al-98} and \cite{Nin:al-08}, and references therein, for a more detailed presentation of radio emission processes.
 
 The data were provided by the following instruments: (a) SXR from the Geostationary Operational Environmental Satellites  (GOES) operated by NOAA\footnote{\url{http://www.ngdc.noaa.gov/stp/SOLAR/}}; (b) flux density time profiles at selected  radio frequencies from the Radio Solar Telescope Network (RSTN; 0.245-15.4~GHz range) of the US Air Force\footnote{provided by NGDC/WDC Boulder \url{http://www.ngdc.noaa.gov/stp/space-weather/solar-data/solar-features/solar-radio/rstn-1-second/}} and the Radio Polarimeters of the Nobeyama Radio Observatory \citep[NoRP; 1.0-80~GHz;][]{Tri:al-79,Nak:al-85}\footnote{\url{http://solar.nro.nao.ac.jp/}}; (c) decimetre-to-metre wave dynamic spectrograms by the ARTEMIS~IV\footnote{Appareil de Routine pour le Traitement et l'Enregistrement Magn\'etique de l'Information Spectrale, \url{http://web.cc.uoa.gr/~artemis/}} solar radio spectrograph at Thermopylae \citep[Greece;][]{Car:al-01,Ktg:al-06,Ktg:al-08}; (d) dynamic spectra at decametric-to kilometric waves (14 MHz to some tens of kHz) by the WAVES spectrograph aboard the Wind spacecraft \citep{Bou:al-95}. The LEAR flux densities at 8.8 and 15.4~GHz suffered from saturation near the peak of the burst, while the NoRP observations ended during the burst, shortly before 7~UT. The flux density time history labelled ``NoRP 17 \& LEAR 15.4~GHz" in  Fig.~\ref{Fig_ovw}.b is a composite of the 17~GHz NoRP profile before and the 15.4~GHz LEAR profile after 06:55~UT.

\subsection{Time profile of relativistic protons detected at Earth}
\label{Sec_Obs_p}

The top panel of Fig.~\ref{Fig_ovw} shows the intensity-time history of relativistic protons after subtraction of a travel time of 216~s. This travel time was evaluated in Paper~1 using the hypothesis that the first relativistic protons escaped to space together with the first electron beams emitting the strong Type~III bursts below 14 MHz (Fig.~\ref{Fig_ovw}.d). A remarkable feature of the time profile is its double-peaked structure. As shown in Paper~1 (Fig.~1), the first peak was nearly exclusively due to anti-Sunward streaming protons, while during the second peak, anti-Sunward streaming protons were observed on top of an isotropic population. 

The question of whether the second rise in the proton profile is a signature of a second solar particle release or not is debated in the literature. Particle reflection at magnetic bottlenecks beyond 1~AU due to previous CMEs or magnetic mirroring in an extended loop has been discussed in some other GLEs \citep{Bie:al-02,Sai:al-08}, and was also proposed for the 2005 Jan 20 event \citep[see Abstract in ][]{Ruf:al-10}. But since the second pulse is due to anti-Sunward streaming protons, its timing cannot be explained by reflection of particles from the first release, for one would have to require that the reflected particles travel back to the Sun and are again reflected Earthward. Even if such a process could  create a new rise, the time needed for the particles to travel Sunward from the bottleneck and again Earthward is too long. From the estimated travel time of 12.5~min between the Sun and the Earth in the first pulse (see Paper~1), to which the travel time between the Earth and the bottleneck and back to the Earth must be added, the protons would need more than 25~min after the first pulse to reach the Earth a second time. But the second rise starts only about 12~min after the first. This suggests that the second rise in the proton time profile is indeed due to a second release of relativistic protons from the Sun. 
 
 When detected on Earth, the particles from this release are, however, superposed on a strong background, to which the reflection at a magnetic barrier beyond 1~AU may well have contributed. We cannot prove this, because the observations do not allow reconstructing the angular distribution of the protons with minute-scale time resolution, which would be needed to determine the time when the isotropic background appeared. However, the contribution of reflected protons after the first peak is plausible in light of other GLEs and of the complex interplanetary magnetic field on Jan 20, in response to CMEs on previous days \citep[see][]{Msn:al-12}. We suggest that the reduced anisotropy during the second release was not a signature of the acceleration process, but rather due to the presence of this background.

In summary we consider that the two pulses in the relativistic proton time profile result from separate acceleration processes near the Sun. The actual time profile of the second pulse depends strongly on the fitted parameters of the rigidity spectrum. This is seen by the comparison of the time profiles at different proton energies in Fig.~\ref{Fig_ovw}.e, with an impulsive peak at 2~GV, at the onset of the long second pulse seen at 5~GV. In the following comparisons, we consider the existence of the new release as established, but its duration as uncertain.

%
\begin{table}[h]
 \caption[]{Acceleration episodes during the 2005 Jan 20 event.}
\label{Tab_eps}
\small
\begin{tabular}{clll}
 \hline \hline
  Episode &
  EM emission &
 GLE signature  \\
 (start time) \\
  \hline
0 ($\sim$06:36)  & Early SXR rise; cm-dm-$\lambda$;          & none observed  \\
       & no m-km-$\lambda$  & \\
1 (06:43:30)	& SXR, HXR, cm-m-$\lambda$ rise;                    & none observed  \\
			& no Dm-km-$\lambda$   &  \\
2-4 (06:45:30) 	& SXR rise; $\pi$-decay $\gamma$,                           & 1st peak  \\
			& main HXR, cm-$\lambda$; & \\
			& dm-km-$\lambda$ (IV, III)   & \\
5 (06:53)	& SXR peak and early decay;               & 2nd peak  \\
		& HXR, mm-short cm-$\lambda$ decay;& \\
      		 & further long cm-km-$\lambda$ (IV, III) & \\
6 (07:06)	& SXR decay,      & decay 2nd peak \\
 		& minor cm-$\lambda$ peaks, m-$\lambda$ IV,                                             & \\
       		& DH$^{\rm (a)}$ II, isolated DH III               & \\
       \hline
  \end{tabular}
\mbox{  (a) ``DH" means dekametric-hectometric. }
\end{table}

   \begin{figure}
   \centering
   \includegraphics[width=8cm]{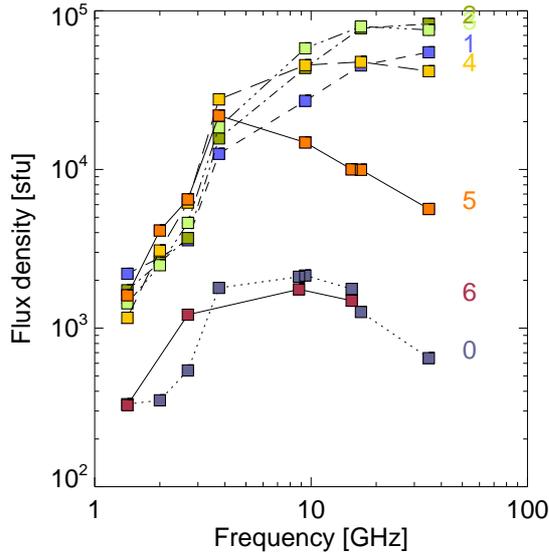}
   \caption[]{Whole Sun radio spectra during the different acceleration episodes, as indicated to the right of the curves. Each spectrum is an average over 1~min taken at the midpoint of the acceleration episode. Data compiled from the Learmonth  RSTN station (1.4, 2.7 GHz during the entire burst, 8.8 GHz during Episodes 0 and 6, and 15.4 GHz during Episodes 0, 5, 6) and the Nobeyama Radio Polarimeters (2, 3.75, 9.4, 17, and 35 GHz, Episodes 0-5). The flux density (ordinate) is given in solar flux units (1~sfu$=10^{-22}$~W~m$^{-2}$~Hz$^{-1}$).
   }
  \label{Fig_gsspec}
   \end{figure}

\subsection{ A tentative identification of acceleration episodes during the event}
\label{Sec_Obs_eps}

The SXR time profile of the event (Fig.~\ref{Fig_ovw}.a) had a smooth rise and fall, while the gamma-ray, HXR, and radio  emission evolved in distinct steps that allow us to roughly distinguish different episodes of particle acceleration. They are labelled from 0 to 6 in Fig.~\ref{Fig_ovw} and are summarised in Table~\ref{Tab_eps}. Episodes 1 to 4 were discussed in Paper 1, and the others will be introduced in Sects.~\ref{Sec_Obs_pr1} and \ref{Sec_Obs_pr2}. The hard X-ray (HXR) emission is not shown in Fig.~\ref{Fig_ovw}. Its overall evolution was similar to the high-frequency ($\sim$9-17~GHz)  microwaves (see Fig.~2 of Paper~1). Microwave spectra during some of the acceleration episodes are plotted in Fig.~\ref{Fig_gsspec} using data from LEAR and NoRP. A check of consistency between the two observatories and the peak flux densities reported in {\it Solar Geophysical Data, comprehensive reports} 731 Part~II (NOAA) resulted in the elimination of the records at 1~GHz from NoRP and at 4.995~GHz from LEAR during all intervals, of the 8.8 and 15.4~GHz data from LEAR during intervals 1-5 and 1-4, respectively. 

In the following we give a short account of the electromagnetic emissions before and during the first proton peak and then focus on the second peak.

\subsection{Before and during the first relativistic proton release - a brief overview}
\label{Sec_Obs_pr1} 
 
The radio emission started increasingly later with decreasing frequency: episode 0 during the early rise of the SXR burst comprised cm-mm wave gyro-synchrotron emission since 06:36~UT, as shown by the flux density spectrum labelled ``0" in Fig.~\ref{Fig_gsspec} (dotted curve). This emission had no counterpart at metric and longer waves \citep[for details see Figs. 1-3 of][]{Brt:al-10}. The low-frequency limit of the radio bursts drifted gradually towards lower frequencies. This well-defined low-frequency cutoff and the absence of Type~III burst emission suggest that the electron acceleration proceeded in low-lying coronal structures -- typically the height range of SXR and microwave emission -- and that the electrons remained confined there during episode~0. The gradual progression towards lower frequencies shows that the confining magnetic structures were evolving or that the acceleration region successively comprised more extended structures. 
 
As shown in Paper~1, the first peak of the relativistic proton profile occurred together with two radiative signatures in Episode~2: with the rise of pion-decay gamma rays, produced by the interaction of protons at energies above 300~MeV with the dense chromosphere \citep[see][and references therein for a more detailed discussion of pion-decay gamma rays from relativistic protons]{Mur:al-87,Vil:al-03a}, and with the first intense Type~III emission below a few tens of MHz (Fig.~\ref{Fig_ovw}.d) signalling electron beams that escape to interplanetary space. The proton acceleration was attributed to processes in the flaring active region, presumably related to magnetic reconnection. The simultaneous occurrence of the gamma rays and the Type~III bursts suggests that the flare-accelerated relativistic protons could escape to the interplanetary space, too, and reach the Earth. The time delay to the start of the GLE then implied an interplanetary path length of about 1.5~AU. It is larger than the nominal Parker spiral, but consistent with a distorted field line in the neighbourhood of an interplanetary CME \citep{Msn:al-12}.

The microwave spectrum (Fig.~\ref{Fig_gsspec}) had a typical gyro-synchrotron shape during the impulsive flare phase (Episodes 1-4), rising up to high frequencies, with a maximum near or above 30~GHz. Optically thick emission up to 30 GHz is consistent with intense magnetic fields close to sunspots, where the optical and UV ribbons were observed during the impulsive phase \citep{Grc:al-08}. These authors evaluated a magnetic field strength of 1600~G.

\section{The second relativistic proton release}
\label{Sec_Obs_pr2}

\subsection{Relativistic proton release and electromagnetic emissions}

The relative timing of relativistic protons on Earth and radio emissions derived for the first peak also determines the relative timing during the second peak. As seen in Fig.~\ref{Fig_ovw}.d, the rise to this second peak, starting near 06:57~UT, was accompanied by a new  group of decametric-to-hectometric (DH) Type~III bursts.  The release of electron beams to the high corona and interplanetary space at this specific time strengthens the conclusion that the second rise of the relativistic proton profile was due to a new solar particle release. A new episode of near-relativistic electron acceleration is also seen as a minor peak superposed on the decaying 15.4-17~GHz time profile in Fig.~\ref{Fig_ovw}.b, and as a clear new peak at 2.7~GHz. Based on these time profiles, we identify new acceleration episodes: Episode~ 5, which includes the onset of the second proton peak, and Episode~6 for later acceleration, which produced weaker new peaks at microwave frequencies. 

The whole Sun radio spectra derived from the fixed-frequency observations still have the gyro-synchrotron shape in Episodes 5 and 6, but with a much lower peak frequency, near 5~GHz, than in the impulsive phase. This indicates radiation from a source with much weaker magnetic field than in the impulsive phase. The gyro-synchrotron spectra extend well below 1~GHz, as suggested by the uniformly grey background that persists in the dm-m wave spectrum throughout Episodes 1 to 6 (Fig.~\ref{Fig_ovw}.c). This shows that near-relativistic electrons accelerated during Episodes~5 and 6 were released into magnetic structures within an extended height range in the corona. 

\begin{figure}
\centering
\includegraphics[width=8cm]{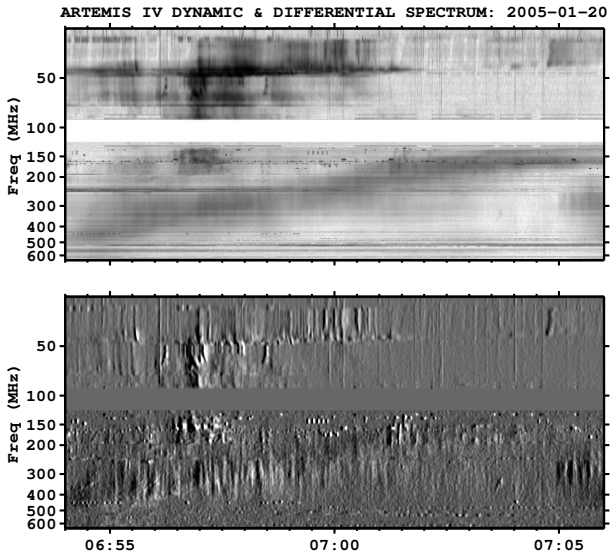}
\caption[]{Detail of the dynamic spectrum (ARTEMIS~IV; top; inverse grey-level scale) and differential spectrum (bottom) during the narrowband drifting lane in Episode 5. The vertical axis gives the frequency in MHz. The series of dark spots near the low-frequency border of the total flux density spectrum is an instrumental artefact.}
\label{Fig_Cont}
\end{figure}

Superposed on the gyro-synchrotron emission during Episode~5 is a band of emission at dm-m wavelengths that crosses the frequency range 650-100~MHz in Fig.~\ref{Fig_ovw}.c. A more detailed view is shown in Fig.~\ref{Fig_Cont}. The emission undergoes a systematic drift from high (06:55~UT near 500~MHz) to low frequencies ($\sim$07:02~UT near 150~MHz). It is accompanied on its low-frequency side (especially 20-80 MHz, around 06:57~UT) by radio emissions that connect to the Type~III burst below 14~MHz, indicating that electrons got access to interplanetary space at least during the early part of Episode~5.

\begin{figure}
\centering
\includegraphics[width=7.5cm]{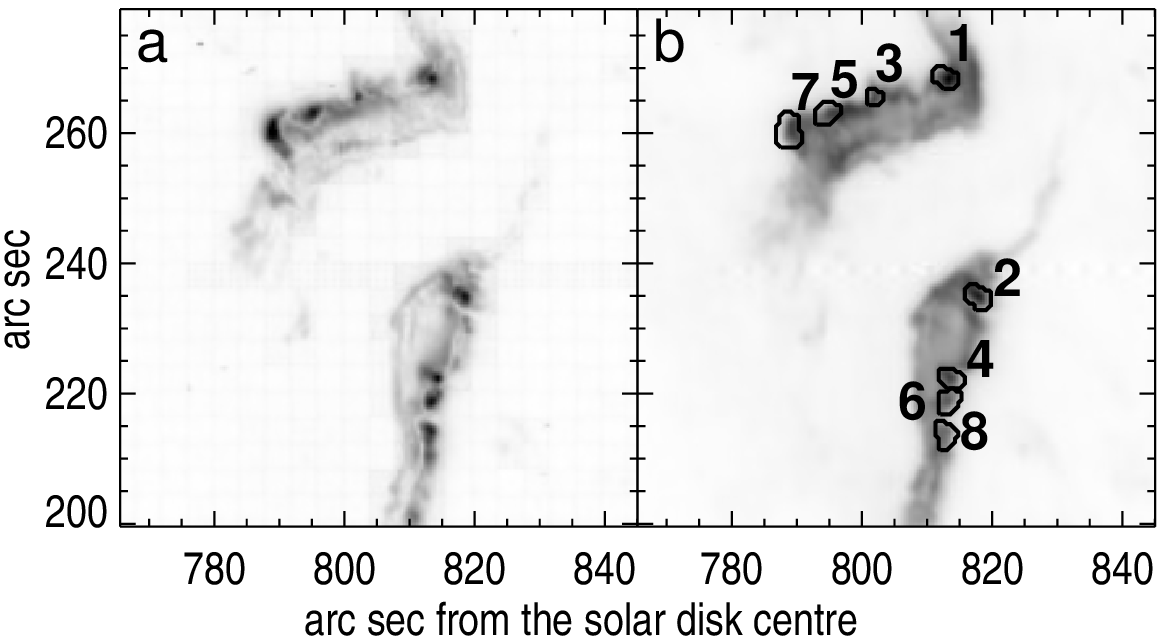}
\includegraphics[width=7.5cm]{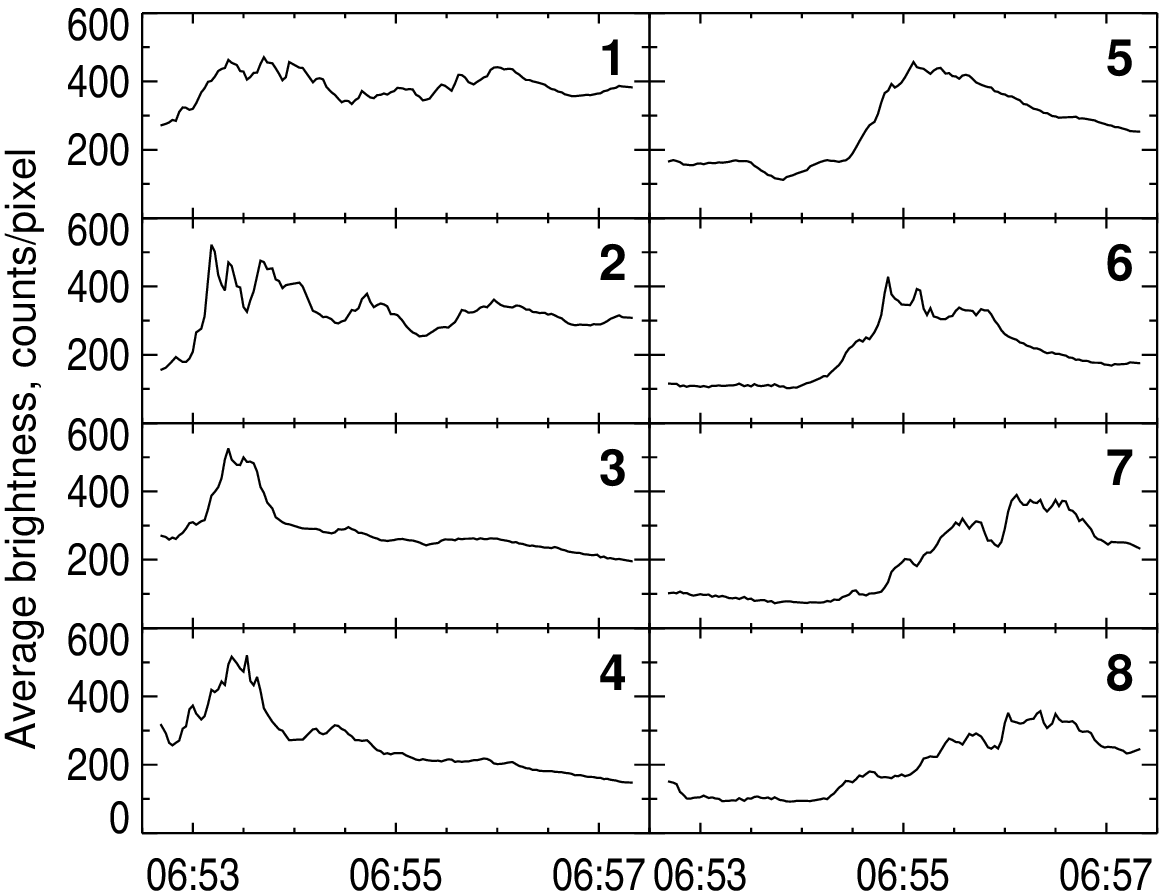}
\caption[]{Flare ribbons and flare kernels (TRACE 160~nm) during the impulsive and early post-impulsive phases of the 2005 Jan 20 flare. 
{\it Top}: maps of the variance of the emission (a) showing the flare kernels and of the average emission (b), showing the flare ribbons, with overlaid flare kernels as identified in the variance map (50\% contour; see text). The axes are labelled in arc seconds from disk centre.
{\it Bottom}: time profiles of the individual kernels, numbered as in map (b). Odd-numbered  kernels are in the northern ribbon, even-numbered ones in the southern ribbon. Successive odd and even numbers denote conjugate kernels.}
\label{Fig_TRACE}
\end{figure}

A clear new episode of energy release starting near 06:54~UT can also be seen in UV observations during Episode~5, and the sources can be localised in flare kernels at 160~nm wavelength with the {\it Transition Region and Coronal Explorer} \citep[TRACE;][]{Hnd:al-99}. An average image of the flaring active region was computed using 20 snapshots between 06:52:41 and 06:57:30~UT, together with a map of the variance of the brightness time history \citep[see][]{Grc-03}. The variance map in the top left-hand panel of  Fig.~\ref{Fig_TRACE} shows the most variable emission in the flare kernels. The 50\% contours of the variance map are overlaid on the average image, which displays the flare ribbons, in the top right-hand  panel. The individual light curves in the bottom panels of Fig.~\ref{Fig_TRACE} display the time histories of brightness in the kernels labelled 1 to 8. Successive pairs show similar light curves from opposite flare ribbons. Their similarity suggests that the kernels are conjugate footpoints of flaring loops. Of particular interest for our present discussion are the light curves of kernels (5, 6) and (7, 8), because they show a clear activation at 6:54~UT. These kernels were located on the outer sides of the average flare ribbons, respectively at the eastern and southern extremes of the northern and southern ribbons. The time profiles show that a new episode of energy release started at 06:54~UT, together with the new microwave peak (Fig.~\ref{Fig_ovw}.b) and the drifting band of dm-m wave radio emission (Figs.~\ref{Fig_ovw}.c, \ref{Fig_Cont}). The energy release starting 06:54~UT thus involved part of the structures that brightened in the impulsive phase (kernels 1, 2) and newly brightening kernels of pairs (5,6) et (7, 8) that probably relate to more extended coronal magnetic structures than the previous episodes, because the kernels are further apart than previously. The first post flare loops became visible between the flare ribbons in TRACE UV images at about 7:04~UT, well after the brightening of these flare kernels, and persisted throughout the day \citep{Grc:al-08}. The kernels (5, 6) and (7,8) thus signal a distinct early energy release process during Episode~5, but the subsequent appearance of post flare loops suggests that energy release continued at increasing coronal height. 

The nature of the slowly drifting band of radio emission in the 500-100~MHz range (Fig.~\ref{Fig_Cont}) can give hints at the electron acceleration process during Episode~5 and, because of the common timing, possibly also hints at the relativistic proton acceleration. Because of its drift rate and relative bandwidth, this feature has been identified as a Type~II burst in several publications \citep{Poh:al-07, Msn:al-09}. \cite{Poh:al-07} recognised that given the frequency range, which is expected to be emitted within a solar radius above the photosphere, the radio source could not be located at the front of the CME that was seen at heliocentric distance 4.5~\RSUN by SoHO/LASCO at 06:54~UT. But it could be due to a shock driven by the lateral expansion of the CME \citep[see, for instance, the coronal shock shape in the simulations of][]{Pom:al-08}. 

However, a more detailed analysis of the ARTEMIS~IV radio spectrum shows that this radio emission is not a Type~II burst. The total and differential dynamic spectrograms of the drifting lane in Fig.~\ref{Fig_Cont} show a drift rate of about $-0.8 \; \rm MHz~s^{-1}$ (relative drift rate $-0.0026 \; \rm s^{-1}$) and an instantaneous bandwidth of about 30\%  of its central  frequency. While these values are typical of Type~II bursts at these frequencies \citep{Man:al-95}, the spectrum is basically a continuum with different types of fine structure during different time intervals. None of the fine structures that are common in Type~II bursts are seen, such as fundamental/harmonic structure, splitting of the band into narrow lanes, or short bursts with rapid frequency drift called herringbone bursts \citep{Nel:Mel-85,Nin:al-08}. 

\begin{figure}
\centering
\includegraphics[width=9cm]{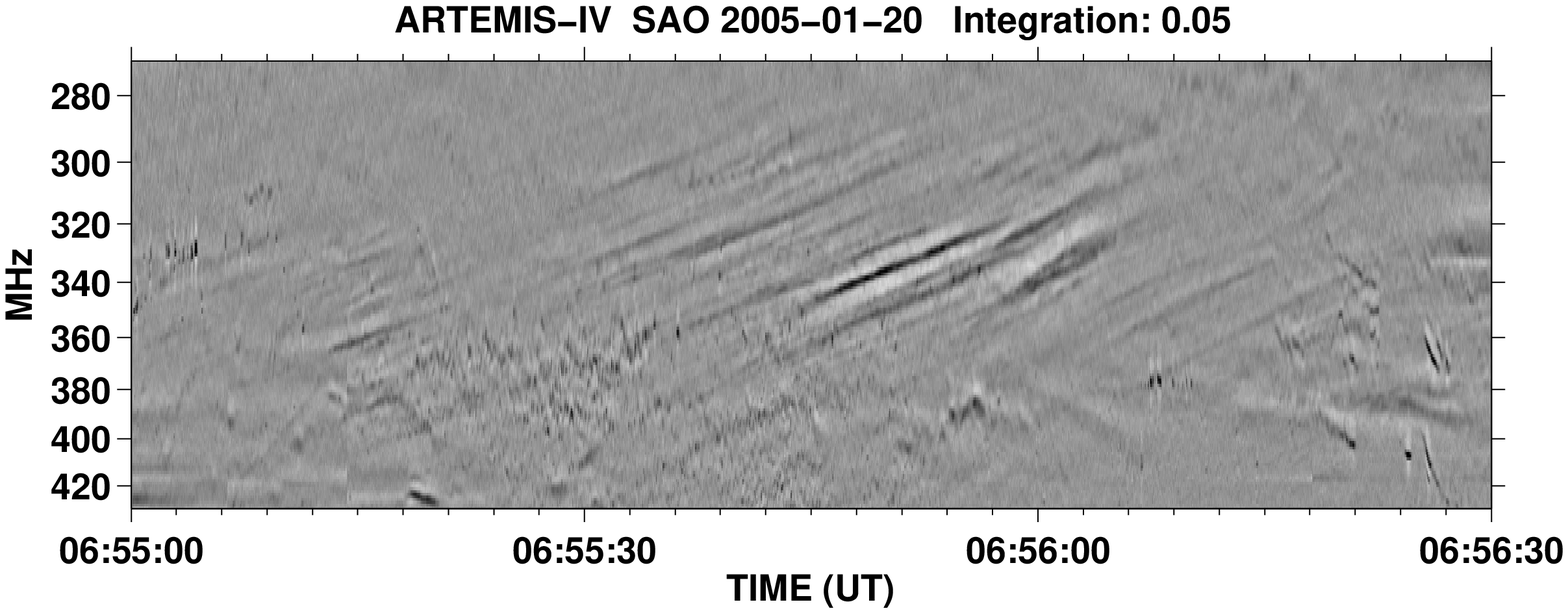}
\includegraphics[width=9cm]{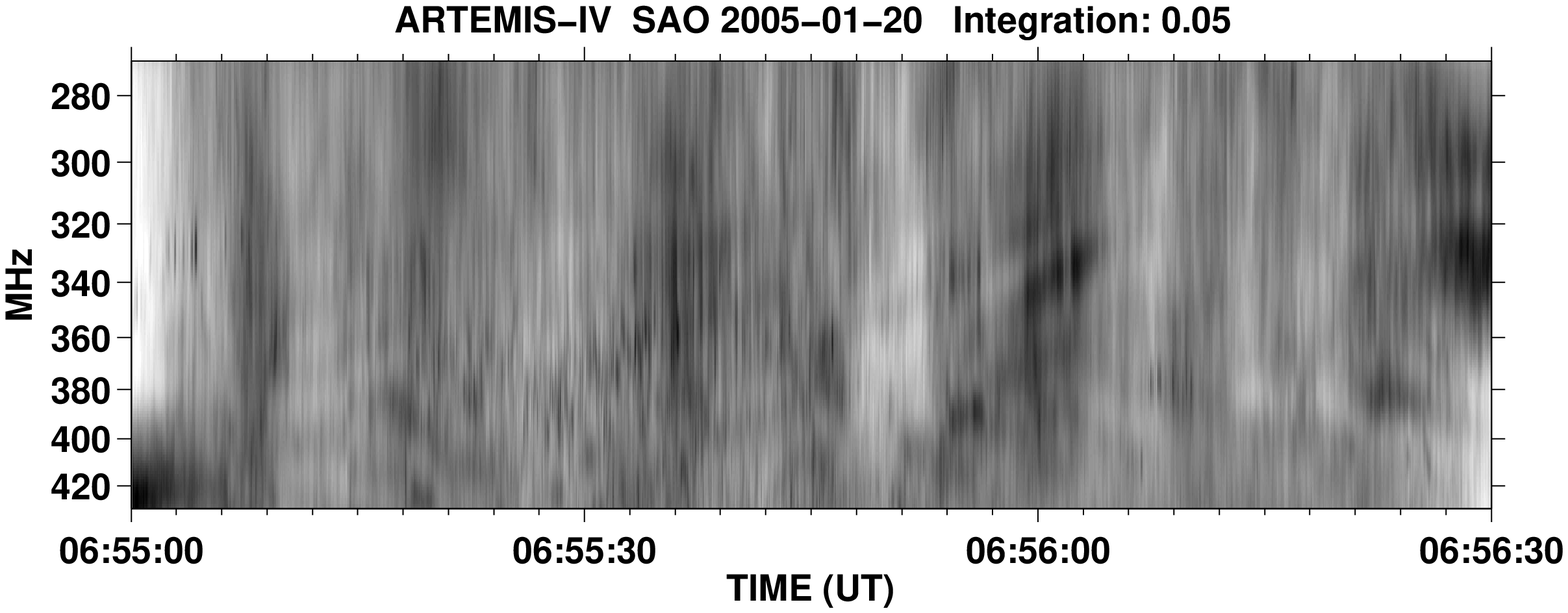}
\caption[]{Details of the dynamic spectrum of the drifting continuum in Fig.~\ref{Fig_Cont} with high spectral resolution: fibre bursts and narrow-band spikes (top), broadband pulsations (bottom).}
\label{Fig_Cont_detail}
\end{figure}

The most prominent fine structures of the drifting lane, which are clearly seen in the differential spectrum (Fig.~\ref{Fig_Cont}, bottom), are broadband pulsations. A detailed spectrum of the early phase is shown in Fig.~\ref{Fig_Cont_detail}, using the high spectral resolution mode of the ARTEMIS~IV acousto-optical receiver (SAO). The high-resolution spectra were high-pass filtered along the time axis to suppress both the continuum background and the terrestrial emitters, which show up as narrow lines parallel to the time axis (see, for instance, the upper panel of Fig.~\ref{Fig_Cont}). Then we separated fibres and spikes from pulsations. To this end we used high-pass filtering along the frequency axis to eliminate pulsations (Fig.~\ref{Fig_Cont_detail}, upper panel), and low-pass filtering to suppress fibre bursts and spikes  (Fig.~\ref{Fig_Cont_detail}, lower panel). Besides the pulsations, which are clearly visible in the lower panel of Fig.~\ref{Fig_Cont_detail}, the drifting emission feature shows narrow-band spikes or zebra pattern structures above about 340~MHz (upper panel) and time-extended fibre bursts with negative frequency drift below 360~MHz. \cite{Chv-11} gives a detailed review of fine structures in solar radio bursts. All spectral features of the slowly drifting lane of dm-m wave emission that accompanied the second relativistic proton release were typical of  a burst of Type~IV, also called a flare continuum. This gives a clue to the acceleration process, as discussed in Sect.~\ref{Sec_Obs_pri}.

The 2005 Jan 20 event was clearly eruptive, with a fast CME \citep{Grc:al-08} and an EUV wave \citep{Mit:al-14}. These phenomena were very likely accompanied by shock waves and therefore Type~II radio bursts. The dynamic radio spectrum below 14~MHz (Fig.~\ref{Fig_ovw}.d)  shows a Type~II burst at its high-frequency border near the start of Episode~6 ($\sim$07:06 in the 10-14~MHz range). This burst is labelled II(3) in \cite{Brt:al-10}. It was not the low-frequency extension of a metre wave Type~II burst - at least no corresponding signature is seen in Fig.~\ref{Fig_ovw}.c. This suggests that the shock-related radio emission truly started in Episode~6. It followed in time the drifting Type~IV burst, and may well be related to the expanding CME structures of which the Type~IV source is part. There is no time correspondence between the Type~II burst and the start of the second relativistic proton peak.

\subsection{An interpretation of particle acceleration during the second relativistic proton release}
\label{Sec_Obs_pri}

Here we attempt an interpretation of the observations during Episode 5, related in particular to the origin of the drifting Type~IV burst. Although we have no imaging observations of the dm-m wave radio sources on 2005 Jan 20, we can refer to well-observed events reported in the literature where a consistent scenario of radio emission during the formation and outward expansion of a flux rope has been developed. The imaging observations typically show the presence of an outward moving outer source and one or more sources at lower altitude with or without systematic motion \citep{Lai:al-00,Kle:Mou-02,Vrs:al-03a,Pic:al-05,Mai:al-07,Hua:al-11,Dml:al-12}. 

Within a classical scenario of flux rope formation, as described by \cite{Dml:al-12}, the outward moving radio source is ascribed to electrons accelerated during reconnection in the underlying current sheet, from where they are injected upwards onto freshly reconnected field lines that are draped around the flux rope. Since the flux rope moves outwards and expands, the internal thermal electron density decreases, and the plasma emission of the confined electrons occurs at gradually decreasing frequencies, as observed in the Type IV continuum during Episode~5. The moving radio source is often observed along the extrapolated trajectory of an erupting prominence \citep{McL-73,Ste:al-78,Mac-80,Kur:al-87,Kle:Mou-02}, so that one expects it to consist of dense plasma well within the CME structure as seen in white light, rather than near its front. Direct comparison of the radio source location with the CME confirms this view \citep[Fig. 11 of ][{\it Erratum}]{Dml:al-12}. This can explain why the Type~IV burst emission on 2005 Jan 20 occurred at rather high frequencies. \cite{Grc:al-14} related a drifting Type~IV burst to the simultaneous expansion of a SXR source. 

Electrons injected downwards from the current sheet generate emission in newly reconnected underlying loops. The near-relativistic part of this population could explain the gyro-synchrotron spectrum with peak frequency near 5~GHz (Fig.~\ref{Fig_gsspec}). Gyro-synchrotron emission from electrons in the overlying flux rope was also reported in other events \citep{Bas:al-01,Mai:al-07,Bai:al-14}, but it can only be detected with imaging observations.

Electron acceleration in the current sheet underneath a CME is in line with other observations:  \cite{Aur:al-09} compare  spectral observations of a Type IV burst having a similar spectrum to the one on 2005 Jan 20 with white-light and spectroscopic UV observations. They conclude that the radio emission came from the current sheet behind a CME. \cite{Ben:al-11} observed the dm sources of two pulsating Type~IV continua above a coronal HXR source in a configuration that again suggested the current sheet of the standard eruptive flare scenario as the acceleration region. \cite{Kli:al-00} presented spectrographic observations of another drifting and pulsating source, which occurred together with the ejection of a soft X-ray plasmoid. They conclude, from a comparison with numerical simulations, that a current sheet underwent repeated reconnection, tearing and coalescence of magnetic islands, leading to repeated bursts of particle acceleration during the gradual build-up of the plasmoid that was eventually ejected

The dm-m-wave radio emission during Episode 5 on 2005 Jan 20 and its simultaneity with the start of the second release of relativistic protons thus suggests that both radio-emitting electrons and relativistic protons were accelerated during the dynamical evolution of coronal magnetic fields in the aftermath of a CME. The scenario involves closed magnetic field lines and does not explain the particle escape to the interplanetary space. But radio emission was also detected on the low-frequency side of the Type~IV burst near the start of Episode 5 (around 06:57~UT, 30-150~MHz in Fig.~\ref{Fig_Cont}). It had a complex spectral structure, as shown in the difference spectrum in the lower panel, with many narrow-band bursts and bursts with positive and negative frequency drift. This was therefore not a typical Type~III burst group, but at still lower frequencies WAVES did see a new Type~III burst (Fig.~\ref{Fig_ovw}.d). It is tempting to interpret the spectral fine structure in the 30-150~MHz range as an indication of magnetic reconnection between the CME and ambient open magnetic field lines \citep[see also][]{Dml:al-12}. This would allow the relativistic protons to escape, as required by our interpretation of the GLE. Such a particle release scenario was modelled by \cite{Msn:al-13}.

We note that the scenario of particle acceleration in the corona behind the CME does not imply that SEP events are always produced after the CME passage. \cite{Kah:al-00} report that events of arcade formation observed in SXR are usually not accompanied by SEP events. Supplementary requirements apparently have to be fulfilled in order to accelerate relativistic protons in this situation, perhaps related to the general energetics of the flaring active region or the CME.

\section{Summary of observational results}
\label{Sec_Obs_sum}

Particle acceleration during the 2005 Jan 20 flare/CME event comprised several successive parts with clear relationships to electromagnetic emissions:
\begin{enumerate}
\item The relativistic proton event observed at Earth (GLE) revealed two distinct solar particle releases.
\item The first relativistic proton release was related in time to signatures of acceleration in and escape from the flaring active region during the impulsive phase: the Type~III emission signalled particle escape to the interplanetary space when pion-decay gamma rays showed the presence of relativistic protons in the low solar atmosphere (see Paper~1).
\item The second relativistic proton release started near the peak of the SXR burst. It was again accompanied by a Type~III burst. A Type~IV burst was observed at dm-m wavelengths. It accompanied new microwave emission, which was more pronounced at frequencies of a few GHz rather than at the higher frequencies that dominated during the impulsive phase acceleration. A clear new energy release was seen in UV flare kernels and was followed by the appearance of post flare loops. The radio and UV emissions are consistent with a scenario of energy release and electron acceleration in the stressed magnetic fields behind the CME, and by inference we suggest that the relativistic protons of the second pulse were also accelerated in this environment.
\item Type~II emission was clearly seen at decametric waves ($\nu \leq 14$~MHz), but started after the second proton peak. Despite the likely presence of shock waves in this very dynamic event, there is no time correspondence between their radio signatures and the acceleration and release of relativistic protons detected at Earth.
\end{enumerate}


\section{Discussion}
\label{Sec_Disc}

The SEP event on 2005 Jan 20 illustrates the long-standing idea, recently substantiated by \cite{McC:al-12}, that GLEs may have a double-peaked structure. One thus expects to see the first peak of the double-peaked structure on 2005 Jan 20 because of the favourable magnetic connection, while the second peak would be visible in all GLEs. This well-defined time structure offers the opportunity, by comparison with electromagnetic emissions, to trace the relativistic protons at 1~AU back to their origin in the corona. The present work uses radio emission, which provides valuable information on electron acceleration and on their release to the interplanetary space, and UV imaging to shed new light on the origin of the late component of the relativistic proton time profile.

The radio emission during the 2005 Jan 20 event shows that non-thermal electrons were injected into successively greater volumes, from compact sources in the early impulsive phase to regions extending to the middle corona after the SXR peak. It is unclear whether nuclear gamma-ray emission was observed after the impulsive phase, because the CORONAS detector has been contaminated by SEPs since the late impulsive phase (Paper~1). Electron acceleration, however, clearly proceeded well beyond the impulsive phase, as shown by the long-lasting radio emission. During the second release of relativistic protons, which started near the maximum of the soft X-ray burst, the hard X-ray and high-frequency ($\geq$8~GHz) microwave emissions were decaying and showed only a minor new peak. But clear rises in emission were found at microwave frequencies below 8~GHz and at dm-m waves, as well as in UV kernels at the periphery of the flare ribbons, with a larger distance between the kernels than before. This is consistent with a new phase of energy release and particle acceleration, very likely in more extended structures and a more tenuous plasma than before. From the properties of the radio emission, we argued that a plausible environment of the acceleration was the post-CME corona with a reconnecting current sheet. 

This scenario goes back to the early flare model of \cite{Car-64}, but can now be substantiated with detailed comparative timing studies. Time-extended acceleration of relativistic protons was discussed with respect to GLEs \citep{Koc:al-94,Aki:al-96,Deb:al-97} and long-duration gamma-ray bursts \citep{Rya-00,Chu:Rya-09}. Time-extended nuclear gamma-ray emission, sometimes persisting over more than ten hours, was discovered by the GAMMA~1 mission \citep{Aki:al-96} and the Compton Gamma-Ray Observatory \citep{Kan:al-93}. The recent FERMI observations with much higher sensitivity have shown that this emission occurs rather often, even in moderate flares \citep{Ack:al-14}. The greater altitude of the acceleration region and the possibly reduced access of the particles to the dense low solar atmosphere -- depending on the magnetic field configuration -- could explain why no such gamma-ray signatures were detected with other GLEs in the past.

Amongst the acceleration mechanisms invoked are direct electric field acceleration in the reconnecting current sheet behind a CME and stochastic acceleration in large-scale turbulent loops \citep[see also][]{Vas:al-06}. Both  are consistent with the association between the proton acceleration and Type~IV radio emission from large-scale loops in the aftermath of a CME. These scenarios show that the CME may be essential for the relativistic proton acceleration even if the acceleration does not occur at a shock wave.

Using diffusive interplanetary transport modelling of the 1991 Jun 15 GLE with the impulsive and post-impulsive microwave time profiles as the injection function, \cite{Aki:al-96} concluded before that post-impulsive acceleration processes were essential to reproducing the neutron monitor observations. Our interpretation of the second peak of the relativistic proton profile during the 2005 Jan 20 GLE is consistent with this and some earlier analyses of relativistic SEP release in temporal association with a Type~IV burst. In the 2000 Jul~14 event \citep[W 07$^\circ$;][]{Kle:al-01} the GLE was accompanied by a Type~IV burst with a spectrum that drifted slowly towards lower frequencies \citep[see the radio spectrum in ][]{Kle:al-01e}. Both the radio spectrum and the relativistic proton timing were similar to Episode 5 of the 2005 Jan 20 event. The timing of the solar release of the first relativistic protons seen at Earth during the 1989 Sep 29 GLE (flare behind the western limb) was also found to be more consistent with a Type~IV burst than with the previous impulsive microwave burst \citep{Kle:al-99a}. In these events we found that the early rise of the GLE was delayed by 10--20~min with respect to the first radiative signatures of electron acceleration in the impulsive phase. Both events were poorly connected. The association of the relativistic proton release with Type~IV emission suggests that a first anisotropic proton pulse was missed in both, in agreement with the  \cite{McC:al-12} scenario. 

\section{Conclusion}

Results for the relativistic SEP event of 2005 Jan 20 presented in Paper~1 and the present work are consistent with the basic scheme of successive relativistic particle releases during a GLE devised by \citeauthor{Mir-01} (\citeyear{Mir-01}, and references therein), \cite{Vas:al-06}, and \citeauthor{McC:al-08} (\citeyear{McC:al-08}, \citeyear{McC:al-12}). Our observations relate the acceleration processes to evolving large-scale magnetic structures in the corona and therefore to the CME development. This does not mean that the CME shock is the accelerator. To the extent that the timing relationship between relativistic protons on Earth and the radio and UV emission in the solar corona discloses a physical relationship, rather than mere coincidence, the observations suggest that relativistic particle acceleration throughout the impulsive and post-impulsive phase is related to the magnetic restructuring of the corona during and after the liftoff of the CME. There is no evidence of a temporal connection between the relativistic proton acceleration and the evolution of a coronal shock as traced by the Type~II radio emission.

\begin{acknowledgements}
The authors are grateful to the Nobeyama Radio Observatory, the CORONAS gamma-ray team, the worldwide network of neutron monitors, and the RSTN network for providing data. KLK acknowledges helpful discussions with J. Bieber, R. B\"utikofer, P. Evenson, E. Fl\"uckiger, D. Lario, V.~ Petrosian, J. Ryan, and K.~Shibasaki, and the kind hospitality of the solar radio astronomy group of the University of Athens and the space research group of the National Observatory of Athens. The referee and the language editor are thanked for most helpful comments on the manuscript. This work was supported in part by the University of Athens Research Center (ELKE/EKPA), by the French Polar Institute (IPEV) and the French space agency (CNES).
\end{acknowledgements}

\end{document}